\documentclass[12pt]{article}
\usepackage{amssymb,latexsym,amsmath}
\begin{document}

\title
{On self-dual Yang--Mills fields\\
in eight and seven dimensions}
\author
{E.K. Loginov\\
Physics Department, Ivanovo State University\\
Ermaka St. 39, Ivanovo, 153025, Russia}
\date{September 26, 2003\footnote{The classification code (MSC)
is 81T13. The keywords are self-dual Yang--Mills
fields.}}\maketitle

\begin{abstract}
The self-duality equations for gauge fields in
pseudoeuclidean spaces of eight and seven dimensions are
considered. Some new classes of solutions of the equations are
found.
\end{abstract}

\section{Introduction}

In 1983 Corrigan et al.~[1] have proposed a generalization of the self-dual Yang--Mills equations in dimension $d>4$:
\begin{equation}
f_{mnps}F^{ps}=\lambda F_{mn},
\end{equation}
where the numerical tensor $f_{mnps}$ is completely antisymmetric and $\lambda=const$ is a non-zero eigenvalue. By the Bianchi identity $D_{[p}F_{mn]}=0$, it follows that any solution of (1) is a solution of the Yang--Mills equations $D_mF_{mn}=0$.
Some of these solutions have found in~[2].
\par
The many-dimensional Yang--Mills equations appear in the
low-energy effective theory of the heterotic string~[3] and in the
many-dimensional theories of supergravity~[4]. In addition, there
is a hope that Higgs fields and super\-symmetry can be understood
through dimensional reduction from $d>4$ dimensions down to
$d=4$~[5].
\par
The paper is organized as follows. Section 2 contains well-known facts about Cayley-Dickson algebras and connected with them Lie algebras. Sections 3 contains the main results.

\section{Cayley-Dickson algebra}

Let us recall that the algebra $A$ satisfying the identities
\begin{equation}
x^{2}y=x(xy),\qquad yx^{2}=(yx)x
\end{equation}
is called alternative. It is obvious that any associative algebra is alternative. The most important example of nonassociative alternative algebra is Cayley-Dickson algebra. Let us recall its construction~(see~[6]).
\par
Let $A$ be an algebra with an involution $x\to\bar x$ over a field $F$ of characteristic $\ne 2$. Given a nonzero $\alpha\in F$ we define a multiplication on the vector space $(A,\alpha)=A\oplus A$ by
\begin{equation}
(x_1,y_1)(x_2,y_2)=(x_1x_2-\alpha \bar y_2y_1,y_2x_1+y_1\bar x_2).
\notag
\end{equation}
This makes $(A,\alpha)$ an algebra over $F$. It is clear that $A$ is isomorphically embedded into $(A,\alpha)$ and $dim(A,\alpha)=2dimA$. Let $e=(0,1)$. Then $e^2=-\alpha$ end $(A,\alpha)=A\oplus Ae$. Given any $z=x+ye$ in $(A,\alpha)$ we suppose $\bar z=\bar x-ye$. Then the mapping $z\to\bar z$ is an involution in $(A,\alpha)$.
\par
Starting with the base field $F$ the Cayley-Dickson construction leads to the following tower of alternative algebras:
\par
1) $F$, the base field.
\par
2) ${\mathbb C}(\alpha)=(F,\alpha)$, a field if $x^2+\alpha$ is the irreducible polynomial over $F$; otherwise, ${\mathbb C}(\alpha)\simeq F\oplus F$.
\par
3) ${\mathbb H}(\alpha,\beta)=({\mathbb C}(\alpha),\beta)$, a generalized quaternion algebra. This algebra is associative but not commutative.
\par
4) ${\mathbb O}(\alpha,\beta,\gamma)=({\mathbb H}(\alpha,\beta),\gamma)$, a Cayley-Dickson algebra. Since this algebra is not associative the Cayley-Dickson construction ends here.
\par
The algebras in 1) -- 4) are called composition. Any of them has the nondegenerate quadratic form (norm) $n(x)=x\bar x$, such that $n(xy)=n(x)n(y)$. In particular, over the field ${\mathbb R}$ of real
numbers, the above construction gives 3 split algebras (e.g., if $\alpha=\beta=\gamma=-1$) and 4 division algebras (if $\alpha=\beta=\gamma=1$): the fields of real ${\mathbb R}$ and complex ${\mathbb C}$ numbers, the algebras of quaternions ${\mathbb H}$ and octonions ${\mathbb O}$, taken with the Euclidean norm $n(x)$. Note also that any simple nonassociative alternative algebra is isomorphic to Cayley-Dickson algebra ${\mathbb O}(\alpha,\beta,\gamma)$.
\smallskip\par
Let $A$ be Cayley-Dickson algebra and $x\in A$. Denote by $R_x$ and $L_x$ the operators of right and left multiplication in $A$
$$
R_x:a\to ax, \qquad L_x:a\to xa.
$$
 It follows from (2) that
\begin{equation}
R_{ab}-R_aR_b=[R_a,L_b]=[L_a,R_b]=L_{ba}-L_aL_b.
\end{equation}
Consider the Lie algebra ${\cal L}(A)$ generated by all operators $R_x$
and $L_x$ in $A$. Choose in ${\cal L}(A)$ the subspaces $R(A)$, $S(A)$, and $D(A)$ generated by the operators $R_{x}$, $S_{x}=R_x+2L_x$, and $2D_{x,y}=[S_{x},S_{y}]+S_{[x,y]}$ respectively. Using (3), it is easy to prove that
\begin{align}
3[R_{x},R_{y}]&=D_{x,y}+S_{[x,y]},\\
[R_{x},S_y]&=R_{[x,y]},\\
[R_{x},D_{y,z}]&=R_{[x,y,z]},\\
[S_{x},S_{y}]&=D_{x,y}-S_{[x,y]},\\
[S_{x},D_{y,z}]&=S_{[x,y,z]},\\
[D_{x,y},D_{z,t}]&=D_{[x,z,t],y}+D_{x,[y,z,t]},
\end{align}
where $[x,y,z]=[x,[y,z]]-[y,[z,x]]-[z,[x,y]]$. It follows from (4)--(9) that the algebra ${\cal L}(A)$ is decomposed in the direct sum
$$
{\cal L}(A)=R(A)\oplus S(A)\oplus D(A)
$$
of the Lie subalgebras $D(A)$, $D(A)\oplus S(A)$ and the vector space $R(A)$.
\par
In particular, if $A$ is a real division algebra, then $D(A)$ and $D(A)\oplus S(A)$ are isomorphic to the compact Lie algebras $g_2$ and $so(7)$ respectively. If $A$ is a real split algebra, then $D(A)$ and $D(A)\oplus S(A)$ are isomorphic to noncompact Lie algebras $g'_2$ and $so(3,4)$.

\section{Solutions of the self-duality equations}

Let  $A$ be a real linear space equipped with a nondenerate symmetric metric $g$ of signature $(8,0)$ or $(4,4)$. Choose the basis \{$1,e_{1},...,e_{7}\}$ in $A$ such that
\begin{equation}
g=\text{diag}(1,-\alpha,-\beta,\alpha\beta,-\gamma,\alpha\gamma,\beta\gamma,-\alpha\beta\gamma),
\end{equation}
where $\alpha,\beta,\gamma=\pm1$. Define the multiplication
\begin{equation}
e_{i}e_{j}=-g_{ij}+c_{ij}{}^{k}e_{k},
\end{equation}
where the structural constants $c_{ijk}=g_{ks}c_{ij}{}^{s}$ are completely antisymmetric and different from 0 only if
$$
c_{123}=c_{145}=c_{167}=c_{246}=c_{275}=c_{374}=c_{365}=1.
$$
The multiplication (11) transform $A$ into a linear algebra. It can easily be checked that the algebra $A\simeq\mathbb O(\alpha,\beta,\gamma)$.
In the basic \{$1,e_{1},...,e_{7}$\} the operators
$$
R_{e_i}=e_{i0}+\frac12c_{i}{}^{jk}e_{jk},\qquad
L_{e_i}=e_{i0}-\frac12c_{i}{}^{jk}e_{jk},
$$
where $e_{ij}$ are generators of Lie algebra ${\cal L}(A)$ satisfying the switching relations
\begin{equation}
[e_{mn},e_{ps}]=g_{mp}e_{ns}-g_{ms}e_{np}-g_{np}e_{ms}+g_{ns}e_{mp}.
\end{equation}
\par
Now, let $H$ and $G$ be matrix Lie groups constructed by the Lie algebras $D(A)$ and  $D(A)\oplus S(A)$ respectively. In the space $A$ equipped with a metric (10) we define the completely antisymmetric $H$-invariant tensor $h_{mnps}$
\begin{align}
h_{ijkl}&=g_{il}g_{jk}-g_{ik}g_{jl}+c_{ijm}c_{kl}{}^{m},
\notag\\
h_{ijk0}&=0,
\notag
\end{align}
and the completely antisymmetric $G$-invariant tensor $f_{mnps}$
\begin{align}
f_{ijkl}=h_{ijkl},\qquad f_{ijk0}=c_{ijk},
\notag
\end{align}
where $i,j,k,l\ne 0$. The tensors $f_{mnps}$ and $h_{mnps}$ satisfy the identities (cp.~[7])
\begin{align}
f_{mnrj}f^{pstj}&=\delta_m^{(p}\delta_n^s\delta_r^{t)}
-\delta_m^{(s}\delta_n^p\delta_r^{t)}+f_{(mn}{}{}^{(ps}\delta^{t)}_r{}_),\\
h_{mnrj}h^{pstj}&=\delta_m^{(p}\delta_n^s\delta_r^{t)}
-\delta_m^{(s}\delta_n^p\delta_r^{t)}+h_{(mn}{}{}^{(ps}\delta^{t)}_r{}_)-c_{mnr}c^{pst},\\
f_{mnij}f_{ps}{}^{ij}&=6(g_{mp}g_{ns}-g_{ms}g_{np})+4f_{mnps},\\
h_{mnij}h_{ps}{}^{ij}&=4(g_{mp}g_{ns}-g_{ms}g_{np})+2h_{mnps}.
\end{align}
where $(pst)$ is cyclic sum for the indexes.
Define the projectors $\tilde f_{mnps}$ and $\tilde h_{mnps}$ of ${\cal L}(A)$ onto the subspaces $D(A)\oplus S(A)$ and $D(A)$ respectively by
\begin{align}
\tilde f_{mnps}&=\frac18(3g_{mp}g_{ns}-3g_{ms}g_{np}-f_{mnps}),
\notag\\
\tilde h_{mnps}&=\frac16(2g_{mp}g_{ns}-2g_{ms}g_{np}-h_{mnps}).
\notag
\end{align}
It follows from (15)--(16) that
\begin{gather}
f_{mnij}\tilde f_{ps}{}^{ij}=-2\tilde f_{mnps},\\
h_{mnij}\tilde h_{ps}{}^{ij}=-2\tilde h_{mnps}.
\end{gather}
It is obvious that the elements
\begin{gather}
\tilde f_{mn}=\tilde f_{mn}{}^{ij}e_{ij},\\
\tilde h_{mn}=\tilde h_{mn}{}^{ij}e_{ij},
\end{gather}
generate the subspaces $D(A)\oplus S(A)$ and $D(A)$ respectively. Using  the identities (12)--(14), we get the switching relations
\begin{align}
[\tilde f_{mn},\tilde f_{ps}]&=\frac34(\tilde f_{m[p}g_{s]n}-\tilde f_{n[p}g_{s]m})-\frac18(f_{mn}{}^{k}{}_{[p}\tilde f_{s]k}-f_{ps}{}^{k}{}_{[m}\tilde f_{n]k}),\\
[\tilde h_{mn},\tilde h_{ps}]&=\frac23(\tilde h_{m[p}g_{s]n}-\tilde h_{n[p}g_{s]m})-\frac16(h_{mn}{}^{k}{}_{[p}\tilde h_{s]k}-h_{ps}{}^{k}{}_{[m}\tilde h_{n]k}).
\end{align}
\par
Now we can find solutions of (1). We choose the ansatzs~(cp.~[2])
\begin{align}
A_m(x)&=\frac43\frac{\tilde f_{mi}x^{i}}{\lambda^2+x_{k}x^{k}},
\notag\\
B_m(x)&=\frac32\frac{\tilde h_{mi}x^{i}}{\lambda^2+x_{k}x^{k}}.
\notag
\end{align}
Using the switching relations (21)--(22), we get
\begin{align}
F_{mn}(x)&=-\frac49\frac{(6\lambda^2+3x_ix^i)\tilde f_{mn}
+8\tilde f_{mni}{}^s\tilde f_{sj}x^ix^j}{(\lambda^2+x_kx^k)^2},
\notag\\
G_{mn}(x)&=-\frac32\frac{(2\lambda^2+x_ix^i)\tilde h_{mn}
+3\tilde h_{mni}{}^s\tilde h_{sj}x^ix^j}{(\lambda^2+x_kx^k)^2}.
\notag
\end{align}
It follows from (17)--(20) that the tensors $F_{mn}$ and $G_{mn}$ are self-dual. If the metric (10) is Euclidean, then we have the well-known solutions of equations (1)~(see.~[2]). If the metric (10) is pseudoeuclidean, then we have new solutions.

\end{document}